\RequirePackage{fix-cm}
\documentclass[twocolumn,epjc3]{svjour3}
\smartqed  
\RequirePackage{graphicx}
\RequirePackage{mathptmx}      

\usepackage{cite}
\usepackage{appendix}
\usepackage{epsfig}
\usepackage{amsmath}
\usepackage{amssymb}
\usepackage{bm}
\usepackage{hyperref}
\usepackage{slashed}
\usepackage[normalem]{ulem} 
\usepackage{subfig}
\allowdisplaybreaks

\newcommand{\be} {\begin{equation}}
\newcommand{\ee} {\end{equation}}
\newcommand{\ba} {\begin{eqnarray}}
\newcommand{\ea} {\end{eqnarray}}

\newcommand{\Lag}{\mathcal{L}}

\newcommand{\GeV}{\text{ GeV}}

\usepackage{color}
\definecolor{darkblue}{cmyk}{1,0.3,0,0.2}
\definecolor{violet}{cmyk}{0,1,0,0.2}
\hypersetup{colorlinks, bookmarksnumbered, citecolor=darkblue, linkcolor=darkblue, pdfstartview=FitH, urlcolor=darkblue, linktocpage}

\usepackage{xspace}
\newcommand{\MadGraph}{{\rmfamily\scshape MadGraph5\_aMC@NLO}\xspace}
\newcommand{\MadLoop}{{\rmfamily\scshape MadLoop}\xspace}
\newcommand{\MadFKS}{{\rmfamily\scshape MadFKS}\xspace}
\newcommand{\Ninja}{{\rmfamily\scshape Ninja}\xspace}
\newcommand{\Sherpa}{{\rmfamily\scshape Sherpa}\xspace}
\newcommand{\OpenLoops}{{\rmfamily\scshape OpenLoops}\xspace}
\newcommand{\SherpaOpenLoops}{{\rmfamily\scshape Sherpa+OpenLoops}\xspace}
\newcommand{\FeynRules}{{\rmfamily\scshape FeynRules}\xspace}
\newcommand{\FeynArts}{{\rmfamily\scshape FeynArts}\xspace}
\newcommand{\NLOCT}{{\rmfamily\scshape NLOCT}\xspace}
\newcommand{\UFO}{{\rmfamily\scshape UFO}\xspace}
\newcommand{\HPO} {{\rmfamily\scshape HiggsPO}\xspace}

\newcommand{\sqrtS} {{\sqrt{S}}}

\journalname{Eur. Phys. J. C}

\begin{document}

\title{Electroweak Higgs production with \HPO at NLO QCD}

\author{Admir Greljo\thanksref{addr1,addr2,addr3}, Gino Isidori\thanksref{addr1}, Jonas M. Lindert\thanksref{addr4}, David Marzocca\thanksref{addr1,addr5}, Hantian Zhang\thanksref{addr1}}

\institute{Physik-Institut, Universit\"at Z\"urich, CH-8057 Z\"urich, Switzerland \label{addr1}
\and
PRISMA Cluster of Excellence and Mainz Institute for Theoretical Physics, Johannes Gutenberg-Universit\"at Mainz, 55099 Mainz, Germany \label{addr2}
\and
Faculty of Science, University of Sarajevo, Zmaja od Bosne 33-35, 71000 Sarajevo, Bosnia and Herzegovina \label{addr3}
\and
Institute for Particle Physics Phenomenology, Department of Physics, Durham University, Durham,
DH1 3LE, UK \label{addr4}
\and
INFN, Sezione di Trieste, and SISSA, Via Bonomea 265, 34136 Trieste, Italy \label{addr5}
}

\maketitle

\begin{abstract}
We present the \HPO UFO model for Monte Carlo event generation of electroweak $VH$ and VBF Higgs production processes at NLO in QCD in the formalism of Higgs pseudo-observables (PO).
We illustrate the use of this tool by studying the QCD corrections, matched to a parton shower, for several benchmark points in the Higgs PO parameter space. We find that, while being sizable and thus important to be considered in realistic experimental analyses, the QCD higher-order corrections largely factorize.
As an additional finding, based on the NLO results, we advocate to consider 2D distributions of the two-jet azimuthal-angle difference and the leading jet $p_T$ for new physics searches in VBF Higgs production.
The \HPO UFO model is publicly available.
\end{abstract}

\section{Introduction}
\label{sec:intro}

The framework of Higgs pseudo-observables (PO) allows to describe in great generality Higgs production~\cite{Greljo:2015sla} and decay~\cite{Gonzalez-Alonso:2014eva} processes in terms of few parameters defined from the properties of the relevant on-shell physical amplitude.\footnote{See Chapter III.1 of Ref.~\cite{deFlorian:2016spz} for a comprehensive and up to date review of the topic.}
In particular, $h \to 4 f$ decays and electroweak Higgs production processes, associated $p p \to Z h$ and $p p \to Wh$ production ($VH$) as well as vector boson fusion (VBF) production $p p \to h j j$, can be completely characterized by a set of on-shell correlation functions $\langle 0 | \mathcal{T} \{ J_f^\mu(x), J_{f^\prime}^\nu(y), h(0) \} | 0 \rangle$, where $J_{f, f^\prime}^\mu$ are the corresponding fermionic currents for each process. Higgs PO are defined directly from the residues of physical poles in these amplitudes.

The Higgs PO decomposition describes the short-distance contribution to the process and, in particular, encompasses any possible deviation due to some heavy new physics. In order to compare with experimental results it is also important to include long-distance contributions due to soft and collinear radiation. In the case of Higgs decays to four leptons the main effect is due to QED radiation. {As it has been shown in Ref.~\cite{Bordone:2015nqa},  to a very good approximation this effect can be 
accommodated for} in terms of a universal (i.e. independent of the short-distance dynamics) radiator function or, equivalently, by QED shower algorithms.
Instead, in the case of electroweak Higgs production at the LHC, where quark currents are involved, QCD corrections are expected to yield the dominant effect. 
The issue of higher order radiative corrections has been extensively studied in the SM~\cite{Han:1992hr,Figy:2003nv,Bolzoni:2010xr,Bolzoni:2011cu,Cacciari:2015jma,Dreyer:2016oyx,Ciccolini:2007jr,Ciccolini:2007ec,Denner:2014cla,
Ferrera:2011bk,Ferrera:2013yga,Ferrera:2014lca,Englert:2013vua,Goncalves:2015mfa,Denner:2011id,Granata:2017iod,
Dittmaier:2011ti,Dittmaier:2012vm,Heinemeyer:2013tqa,deFlorian:2016spz}, {and} in the context of effective field theory approaches~\cite{Maltoni:2013sma,Mimasu:2015nqa,Degrande:2016dqg}. {The purpose of the present paper is to illustrate how this effect can be accounted for within the PO formalism}.

In order to utilise the Higgs PO formalism in experimental analyses, it is important to embed it into {an appropriate 
Monte Carlo event generator framework. The final goal is the possibility to simulate the electroweak Higgs production processes in a realistic environment,
for arbitrary values of the Higgs PO, and with a theoretical precision exceeding the experimental one}. Such an implementation was initiated in~\cite{Greljo:2015sla}
with the realisation of the universal \FeynRules (\UFO) model, called \HPO, which can be used for example within the \MadGraph~\cite{Alwall:2014hca} or \Sherpa~\cite{Hoche:2014kca} event generators at leading order (LO) in QCD.  {In this paper we present the upgrade} able to generate electroweak Higgs production events at next-to-leading order (NLO) accuracy in QCD matched to parton showers~\cite{webpage}. We present numerical results obtained within \MadGraph. A corresponding interface within \SherpaOpenLoops is under development. The \HPO NLO UFO model is publicly available at\\
\centerline{\url{http://www.physik.uzh.ch/data/HiggsPO}\,\,.}

The paper is organized as follows. In section~\ref{sec:theo} we briefly review the Higgs PO framework for electroweak Higgs production, its dressing with NLO QCD corrections and the corresponding implementation within a UFO model. In section~\ref{sec:results} we illustrate the effect of the NLO QCD corrections in a few relevant kinematical distributions for several benchmark points for both $VH$ and VBF production. In section~\ref{sec:STXS} we use this tool to obtain the expression for the simplified template cross-section bins for $Zh$ production as a function of the Higgs PO at NLO in QCD. We finally conclude in section~\ref{conclusions}.

\section{Theoretical considerations and UFO implementation}
\label{sec:theo}

\subsection{Effective PO Lagrangian}

The PO decomposition for electroweak Higgs production is based on a momentum expansion of the relevant on-shell amplitudes around the physical poles corresponding to the propagation of SM gauge bosons.
In particular, the Higgs PO are defined from the residues of such poles \cite{Gonzalez-Alonso:2014eva,Greljo:2015sla}.
The widely-used \UFO formalism to interface model information with Monte Carlo event generators, instead, relies on amplitude generation starting from a set of interaction Feynman rules, derived from a Lagrangian. 
While the PO are strictly defined from gauge-invariant analytic properties of on-shell scattering amplitudes, the resulting amplitude decomposition can effectively be reproduced by the tree-level matrix elements derived from the following effective Lagrangian~\cite{Greljo:2015sla,Gonzalez-Alonso:2014eva},
\ba
	\Lag^{\rm eff}_{\rm HPO} &=&  {\kappa_{ZZ} }\frac{m_Z^2}{v} Z_\mu Z^\mu h +  {\kappa_{WW} }\frac{2 m_W^2}{v} W_\mu^+ W^{- \, \mu} h + \nonumber\\
	&& - {\epsilon_{\gamma\gamma}}   \frac{h}{ 2v} A_{\mu\nu} A^{\mu\nu} -  {\epsilon_{Z\gamma} }\frac{h}{v} Z_{\mu\nu} A^{\mu\nu} - {\epsilon_{ZZ}} \frac{h}{2v} Z_{\mu\nu} Z^{\mu\nu} \nonumber\\
	&& -  {\epsilon_{\gamma\gamma}^{\rm CP}} \frac{h}{2v} A_{\mu\nu} \widetilde A^{\mu\nu}   -  {\epsilon_{Z\gamma}^{\rm CP}} \frac{h}{v} Z_{\mu\nu} \widetilde A^{\mu\nu} -  {\epsilon_{ZZ}^{\rm CP}}  \frac{h}{2v} Z_{\mu\nu} \widetilde Z^{\mu\nu}  + \nonumber\\
	&& -  {\epsilon_{WW}} \frac{h}{v} W^+_{\mu\nu} W^{- \, \mu\nu} -  {\epsilon_{WW}^{\rm CP}} \frac{h}{v} W^+_{\mu\nu} \widetilde W^{- \, \mu\nu} + \label{eq:HPOeff} \\
	&& + \sum_{f}  \sum_i {\epsilon_{Z f^i } }~ \frac{2 h}{v} Z_\mu \bar{f}^i \gamma^\mu f^i ~ + \nonumber\\
	&& + \sum_{i, j} \frac{2 h}{v} 
	\left[ {\epsilon_{W e^{i}\nu^{j}} } ~ W_\mu^+ \bar{\nu}_{e^i L} \gamma^\mu e^j_L +\right. \nonumber\\
	&& \left. +{\epsilon_{W u^i_L d^j_L}} ~ W_\mu^+ \bar{u}^i_{L} \gamma^\mu d^j_L  + { \epsilon_{W u^i_R d^j_R} }~ W_\mu^+ \bar{u}^i_{R} \gamma^\mu d^j_R 
	+ {\rm h.c.} \right]~,\nonumber
\ea
where $f = \{e_L, e_R, \nu_L, u_L, u_R, d_L, d_R\}$ and $i,j= \{1, 2, 3\}$ are flavour indices.
 Moreover $V_{\mu\nu} = \partial_\mu V_\nu - \partial_\nu V_\mu$ and $\widetilde V^{\mu\nu} =  \frac{1}{2} \epsilon^{\mu\nu\rho\sigma} V_{\rho \sigma}$. The coefficients $\kappa_{VV}, \epsilon_{VV'},  \epsilon^{\rm CP}_{VV'}, \epsilon_{Z,f^i}, \epsilon_{W,f^if^j}$ are considered as the proper Higgs pseudo-observables.
The fla\-vour-universal PO ($\kappa_{VV}$, $\epsilon_{VV'}$, and $\epsilon^{\rm CP}_{VV'}$) 
and the leptonic contact terms ($\epsilon_{Z,e^i}$, $\epsilon_{Z,\nu^i}$, and $\epsilon_{W,e^i\nu^j}$), will most likely be strongly constrained, or measured, from $H\to 4\ell, 2\ell2\nu$ decays.
The remaining contact terms for light quarks can instead be probed
in electroweak Higgs production processes~\cite{Greljo:2015sla}.
The charged-current PO $\epsilon_{W,f_if_j}$ are complex, while all the others are real (in the limit where we neglect re-scattering effects due to light-quark loops). We introduce the complex phase as:
\be
	\epsilon_{W,f_if_j} \to \epsilon_{W,f_if_j} e^{i \phi_{W,f_i f_j}}~,
\ee
where now $\epsilon_{W,f_if_j}$ is taken real.
Assuming the Higgs to be a parity-even state and CP to be conserved, then all couplings are real and all the $\epsilon^{\rm CP}_{VV'}$ vanish.
This Lagrangian, together with the following parametrization of $Vf\bar f$ interactions of the $W$ and $Z$ bosons with fermions,
\be \begin{split}
	\Lag^{\rm eff}_{Vff} &= \sum_{f}  \sum_i  {g_Z^{f_i}}  ~ Z_\mu \bar{f}^i \gamma^\mu f^i  ~+ \\
& 	+ \sum_{i,j} \left[ {(g_{W}^{e_L})_{ij}}   ~W_\mu^+ \bar{\nu}_{e^i L} \gamma^\mu e^j_L 
         +  {(g_{W}^{q_L})_{ij}} ~ W_\mu^+ \bar{u}^i_{L} \gamma^\mu d^j_L  +\right. \\ 
&	\left. +  {(g_{W}^{q_R})_{ij}}  ~ W_\mu^+ \bar{u}^i_{R} \gamma^\mu d^j_R 
	+ {\rm h.c.} \right]~,
	\label{eq:ZWPOeff}
\end{split} 
\ee
and combined with the corresponding gauge boson kinetic terms (based on an unbroken $SU(3)_{\rm QCD} \times U(1)_{\rm QED}$) and ghost terms for QCD, is implemented in a \FeynRules (version 2.3.24)~\cite{Alloul:2013bka} model. 
All SM particle fields are defined in the mass (unitary) eigenbasis with SM QED and QCD gauge interactions. The masses for $W$, $Z$, $h$, and third generation charged fermions are put by hand, while the first two generations and neutrinos are kept massless. The couplings in Eq.~\eqref{eq:HPOeff} are defined with interaction-order label {\tt HPO}.
The corresponding leading-order UFO model generated with \FeynRules is publicly available~\cite{webpage} since the publication of Ref.~\cite{Greljo:2015sla}.

Under the assumption of an $U(2)^3$ flavour symmetry acting on the first two generations of quarks, the number of independent light-quark PO reduces to six \cite{Greljo:2015sla}, namely
\be
\epsilon_{Z u_L},~\epsilon_{Z u_R},~ \epsilon_{Z d_L},~ \epsilon_{Z d_R},~ \epsilon_{W u_L}~, \phi_{W u_L}~,
\label{eq:MFV_contterms}
\ee
or five if we further neglect CP-violating contributions (in which case the phase $\phi_{W u_L}$ vanishes). These six PO are directly accessible in the UFO model (and thus for example in the \MadGraph~{\tt param\_card.dat}) as:
 \begin{equation}
\text{\tt eZuL},~\text{\tt eZuR},~\text{\tt eZdL},~\text{\tt eZdR},~\text{\tt eWuL},~\text{\tt phiWuL}\,.
\end{equation}
This $U(2)^3$ symmetry assumption can be justified by flavour constraints in the light quark sector, as well as by the experimental difficulties of differentiating the light quark flavour. As demonstrated in Ref.~\cite{Greljo:2015sla}, it will be instead possible to separately constrain the PO appearing in Eq.~\eqref{eq:MFV_contterms} by studying in detail $VH$ and VBF Higgs production.

Besides the six quark contact-terms PO listed above, all flavour-universal PO, as well as the leptonic contact terms, the $Zff$ and $Wff'$ vertices, are also implemented. This allows the \HPO model to generate any electroweak Higgs production process, as well as $h \to 4f, ~ 2f\gamma,~Z\gamma,$ and $\gamma\gamma$ decays in the Higgs PO framework at NLO in QCD.

\subsection{Dressing the Higgs PO with NLO QCD corrections}
\label{sec:implementation}

The Higgs PO Lagrangian in Eq.~\eqref{eq:HPOeff} can be considered as an extension
of the EW sector of the SM without new additional states. Therefore, the dressing with 
NLO QCD effects and the matching with QCD patron showers is straightforward based on modern
Monte Carlo frameworks, that include the automated generation of the required 
born, one-loop and real matrix elements, and subtraction of infrared singularities. 
In fact, the HPO effective couplings in Eq.~\eqref{eq:HPOeff} do not require 
UV renormalization at NLO QCD. Still, the inclusion of the quark contact-term PO at NLO QCD 
requires dedicated so-called $R_2$ rational terms.\footnote{
These $R_2$ contributions can be understood as the missing $(4 - D)$-dimensional 
contributions, by construction not necessarily automatically generated within numerical one-loop generators.
These contributions are universal and can be restored from process-independent effective counterterms~\cite{Ossola:2008xq,Binoth:2006hk,Bredenstein:2008zb}. 
}
The contact-term $R_2$ can easily be derived from the $R_2$ contributions 
of the related $Vf_i\bar f_j$ interactions in the SM~\cite{Draggiotis:2009yb}:
\be\begin{split}
	R_2^{\bar{f}_i f_i  Z h} &= -\frac{i g_s^2}{3 \pi^2 v} \epsilon_{Z,f^i}~, \\
	R_2^{\bar{u}_i d_j  W^+ h} &= -\frac{i g_s^2}{3 \pi^2 v} \epsilon_{W,u_L^i d_L^j} e^{-i \phi_{W u}}~.
	\label{eq:R2terms}
\end{split}\ee

Technically, we employ the \NLOCT package (version 1.02)
~\cite{Degrande:2014vpa} together with \FeynArts (version
3.9)~\cite{Hahn:2000kx} to generate all UV and $R_2$ QCD counterterms for the SM
interactions. The resulting model is exported in the UFO format. Since the
present public version of the \NLOCT package is restricted to renormalizable
interactions, we supplement the $R_2$ rational terms related to the contact-term
PO shown in Eq.~\eqref{eq:R2terms} by hand. As already mentioned, UV counterterms for any
of the PO interactions are not needed.

The resulting NLO UFO model can directly be imported into \MadGraph, where all
required one-loop amplitudes are automatically generated with
\MadLoop \cite{Hirschi:2011pa} and \Ninja~\cite{Peraro:2014cba,Hirschi:2016mdz}.
Infrared subtraction of the real contributions is automatically performed \'a la FKS~\cite{Frixione:1995ms} in
\MadFKS \cite{Frederix:2009yq}, where the corresponding real amplitudes are generated from the
underlying UFO model. 
For the VBF Higgs production process the colour suppressed and thus numerically very small 
pentabox contribution in the virtual amplitude can either explicitly be included or excluded (default in \MadGraph). 
The letter case results in a formal mismatch of the IR poles of the virtual and real contributions, which can be dealt with as described in section~\ref{sec:VBF}.
At the same order of perturbation theory as VBF Higgs production also Higgsstrahlung with  
hadronic decays, i.e. $pp\to(V\to q\bar q)H$, contributes to the same $H+2~$jet signature.
Once VBF selection cuts (large invariant masses of the leading jets and/or 
large rapidity separation of the leading jets) are applied, these contributions (and their interference with VBF topologies) are very small. For simplicity in the simulations presented in section~\ref{sec:VBF} we disabled those contributions.
In case the Higgsstrahlung subprocess is not considered as a dedicated background in a VBF analysis, 
it should be included in the VBF process, resulting in additional PO contributions that are automatically generated.
We employ the described implementation in \MadGraph for the numerical predictions
of electroweak Higgs production processes (VBF and $VH$) at NLO QCD matched to
Pythia 6~\cite{Sjostrand:2006za} as presented in the following section.

As a first validation of the code we compare results for the total decay
widths $h \to V \bar{f} f'$ obtained with the \HPO UFO model in \MadGraph at LO
and NLO against a LO analytical computation of the decay width in the PO
formalism. Moreover, the ratio $K$ of the NLO result to the LO one is expected
to be a simple universal factor $K = 1 + \alpha_s / \pi \simeq 1.038$.
For all final states and combinations of contact terms we find perfect
agreement between the analytical and Monte Carlo results within the numerical uncertainties.

As a further cross check and for future applications within other Monte Carlo frameworks, 
we implemented the HPO Lagrangian, as given in Eq.~\eqref{eq:HPOeff},
together with the $R_2$ contributions of Eq.~\eqref{eq:R2terms} for 
the contact-terms in \OpenLoops~\cite{Cascioli:2011va,hepforge}. We compared
the amplitudes for the Higgsstrahlung and VBF Higgs processes at the amplitude 
level and found perfect agreement. Here, for the latter we included the $pp\to(V\to q\bar q)h$ subprocess.
Furthermore, we compared NLO fixed-order differential cross sections 
obtained with \linebreak \MadGraph against \SherpaOpenLoops and found agreement within 
the statistical uncertainty of the Monte Carlo integrations.

\section{Results and Examples}
\label{sec:results}

In order to illustrate the usage of \HPO and the effect of NLO QCD corrections, we {present the simulation of} EW Higgs production processes ($Zh$, $Wh$, and VBF) at the LHC with $\sqrtS=13~$TeV for several benchmark points close to the SM, shown in Table~\ref{tab:bench}.
In each benchmark point a different contact term is switched on, together with the SM contribution ($k_{ZZ} = k_{WW} = 1$).
As parton distribution functions (PDFs) we employ NNPDF23NLO and use the value of $\alpha_S$ they provide.
SM input parameters are chosen in accordance with the defaults of the \HPO UFO model. Renormalisation and factorisation scales for both processes are chosen as $\mu=\mu_{\rm R}=\mu_{\rm F}=H_{\rm T}/2$, where $H_{\rm T}$ is the scalar sum of the $p_{\rm T}$ of all final state particles.

We shower Les Houches events with PYTHIA6 \cite{Sjostrand:2006za} and reconstruct jets using the FastJet \cite{Cacciari:2011ma} implementation of the anti-$k_T$ jet algorithm with $R=0.4$ and a minimum $p_{\rm T,  jet}$ of 10~GeV.

\begin{table}[tb]
\centering
{\footnotesize
\begin{tabular}{lllllll}
BP & \text{\tt eZuL} & \text{\tt eZuR} & \text{\tt eZdL} & \text{\tt eZdR} & \text{\tt eWuL} & \text{\tt phiWuL} \\ \hline
I    & 0 & 0 & 0 & 0 & 0 & 0\\
II   & 0.0195 & 0 & 0 & 0 & 0 & 0 \\
III  & 0 & 0.0195 & 0 & 0 & 0 & 0\\
IV & 0 & 0 & 0.0244 & 0 & 0 & 0\\
V  & 0 & 0 & 0 & 0.0244 & 0 & 0\\
VI  & 0 & 0 & 0 & 0 & 0.018 & 0\\
VII  & 0 & 0 & 0 & 0 & 0.018 & $\frac{\pi}{2}$\\
\end{tabular}}
\caption{Benchmark points in the PO parameter space used to generate events in $p p \to Z H$ (I, II, III, IV and V), $p p \to W H$ (I, VI and VII), and VBF (I, \dots, VII) processes. In all benchmarks we set {\tt kZZ} = {\tt kWW} = 1.} 
\label{tab:bench}
\end{table}

\subsection{Associated $VH$ production}
\label{sec:VH}

\begin{figure}[tb]
  \centering
   \includegraphics[width=0.5 \textwidth]{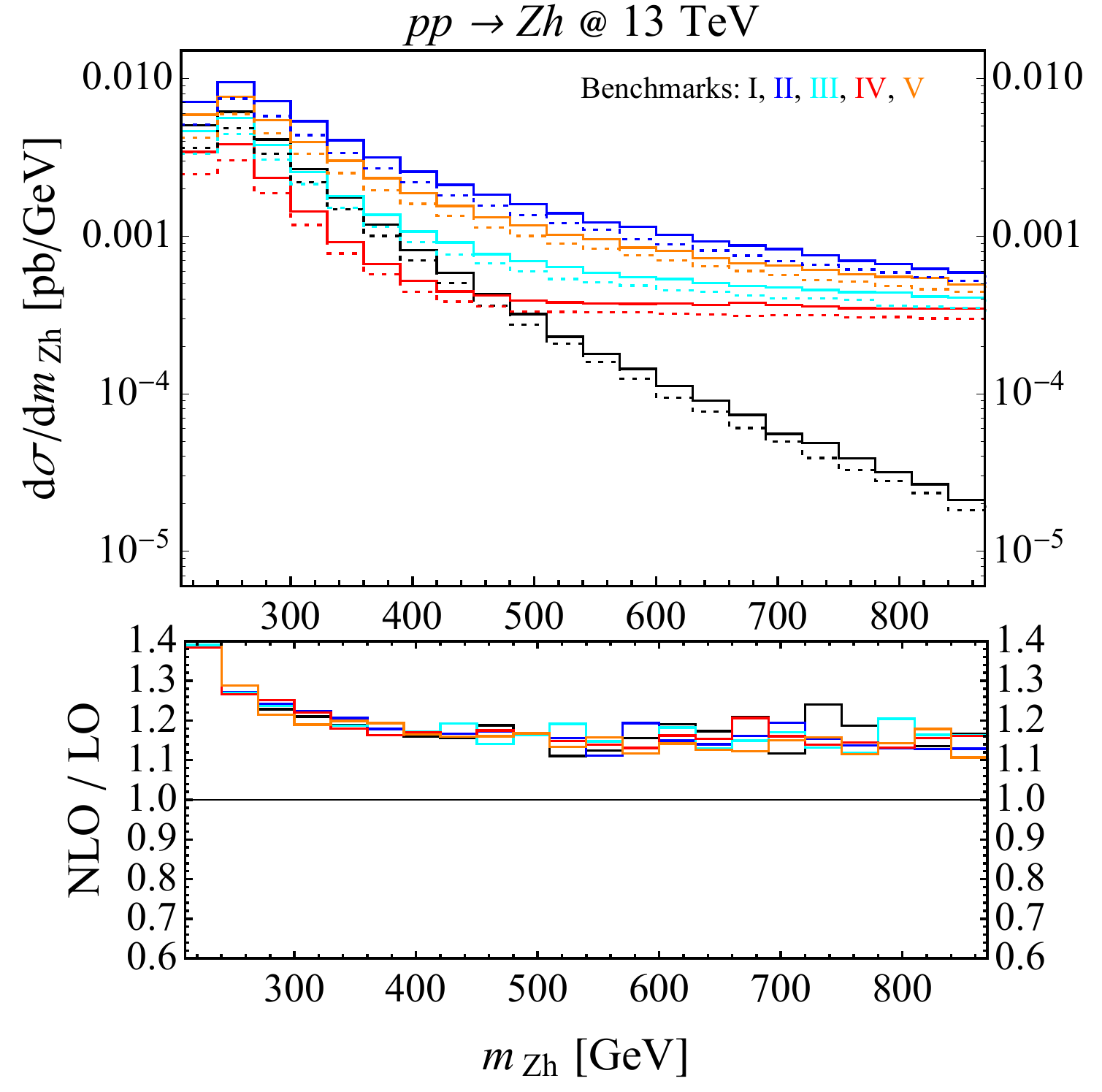}  \\
    \caption{Invariant mass distribution of the $Z H$ system in $p p \to Z H$ production at 13~TeV. Solid (dotted) lines show NLO (LO) + PS predictions for five benchmark points in the PO parameter space listed in Table~\ref{tab:bench}.
    \label{ZH-plot}}
\end{figure}

\begin{figure}[tb]
  \centering
   \includegraphics[width=0.5 \textwidth]{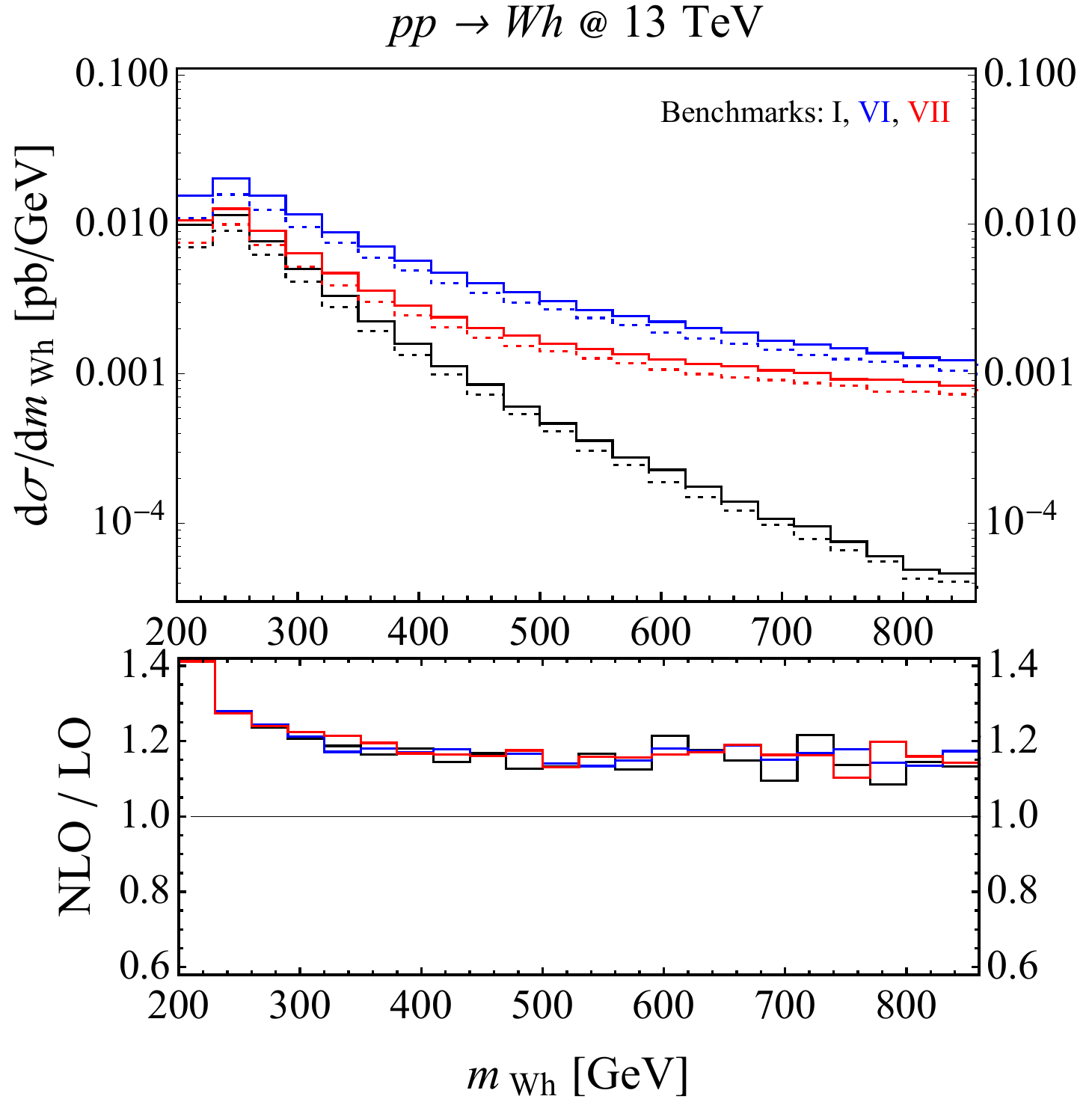}  \\
    \caption{Invariant mass distribution of the $W H$ system in $p p \to W H$ production at 13~TeV. Solid (dotted) lines show NLO (LO) + PS predictions for three benchmark points in the PO parameter space listed in Table~\ref{tab:bench}.
    \label{WH-plot}}
\end{figure}

Events for on-shell $Zh$ and $Wh$ production in the \HPO \linebreak \UFO model can be generated at NLO in QCD with \MadGraph with the following commands\footnote{For generating LO events it is sufficient to remove the {\tt [QCD]} flag.}:
\begin{verbatim}
./bin/mg5_aMC
> import model HPO_ewk_prod_NLO
> generate  p p > z h HPO=1 QED=1 [QCD]
> output
> launch
\end{verbatim}

In Figs.~\ref{ZH-plot} and \ref{WH-plot} we show the $ZH$ and $WH$ invariant mass distributions respectively obtained at NLO (solid lines) and LO (dashed lines) for the benchmark points listed in Table~\ref{tab:bench}. The lower panel of each figure shows the ratio between the NLO and LO results. We observe that, with the choice of $H_{\rm T}$ as dynamical renormalisation and factorisation scales, the NLO $K$-factor is fairly flat and universal in most of the considered kinematic regime for all considered benchmark points.

\subsection{VBF Higgs production}
\label{sec:VBF}

\begin{figure}[tb]
  \centering
   \includegraphics[width=0.49 \textwidth]{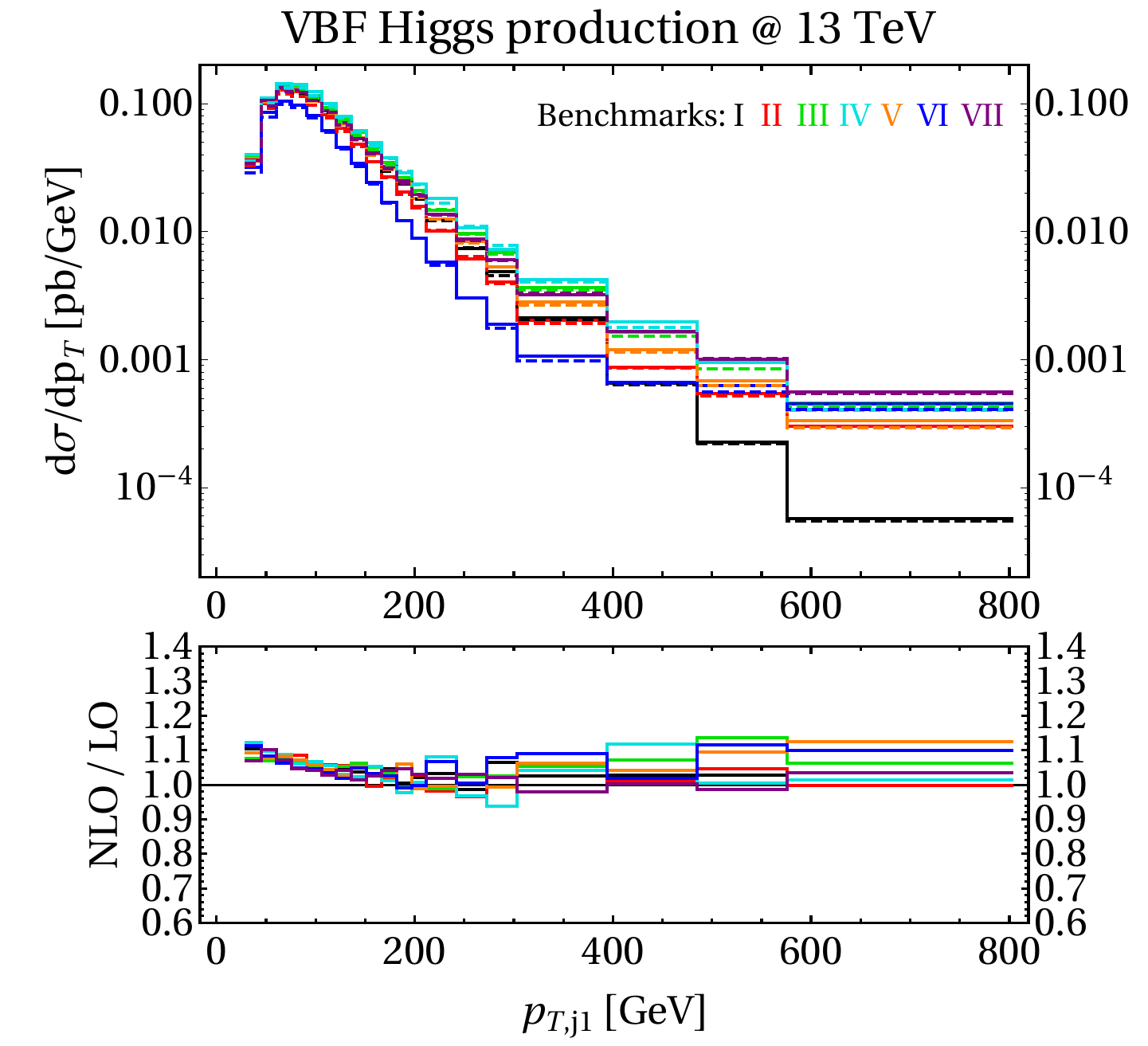}  \\
    \caption{Distribution of the leading-jet $p_T$ in VBF Higgs prodution events at 13~TeV. Solid (dotted) lines show NLO (LO) + PS predictions for seven benchmark points in the PO parameter space listed in Table~\ref{tab:bench}.
    \label{VBF-plot}}
\end{figure}

\begin{figure}[hbt!]
  \centering
  \subfloat{\includegraphics[width=0.4\textwidth]{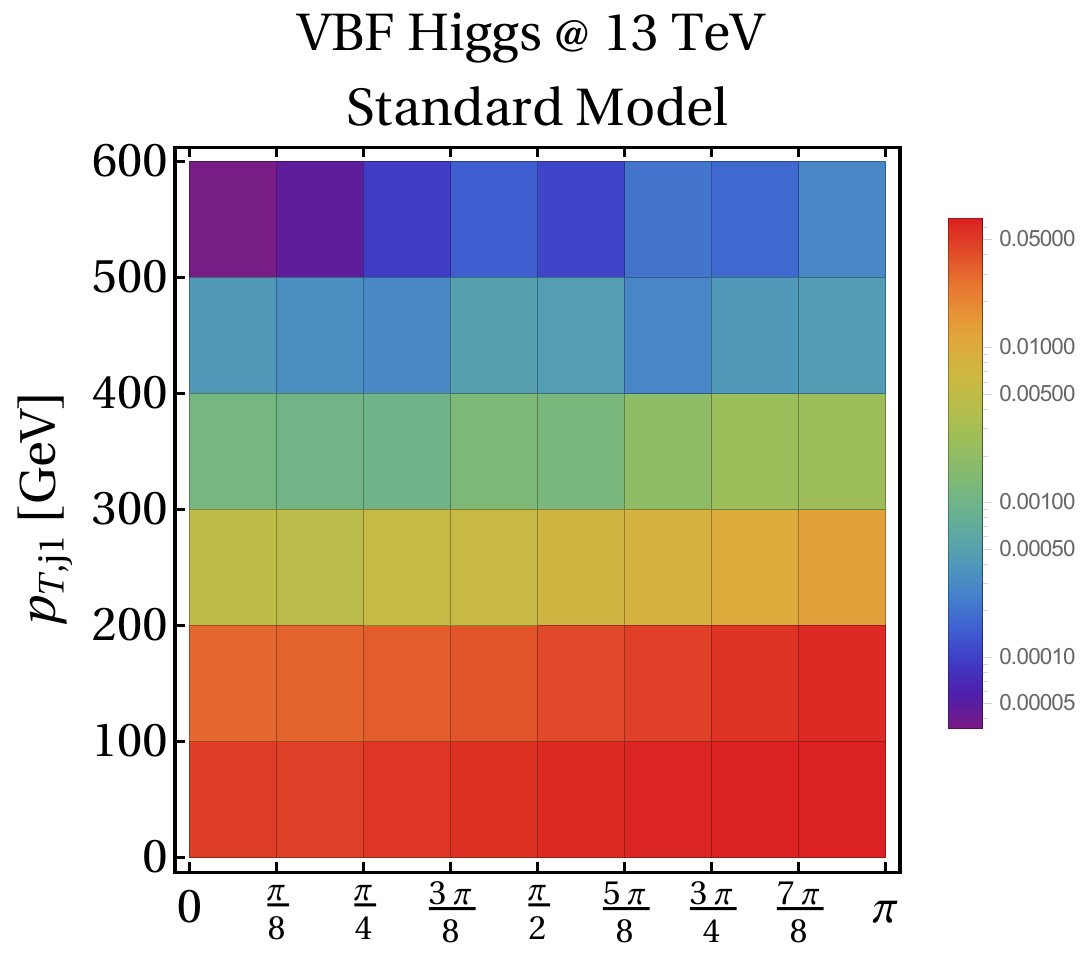}}\\ \vspace{-0.48cm}
  \subfloat{\includegraphics[width=0.4\textwidth]{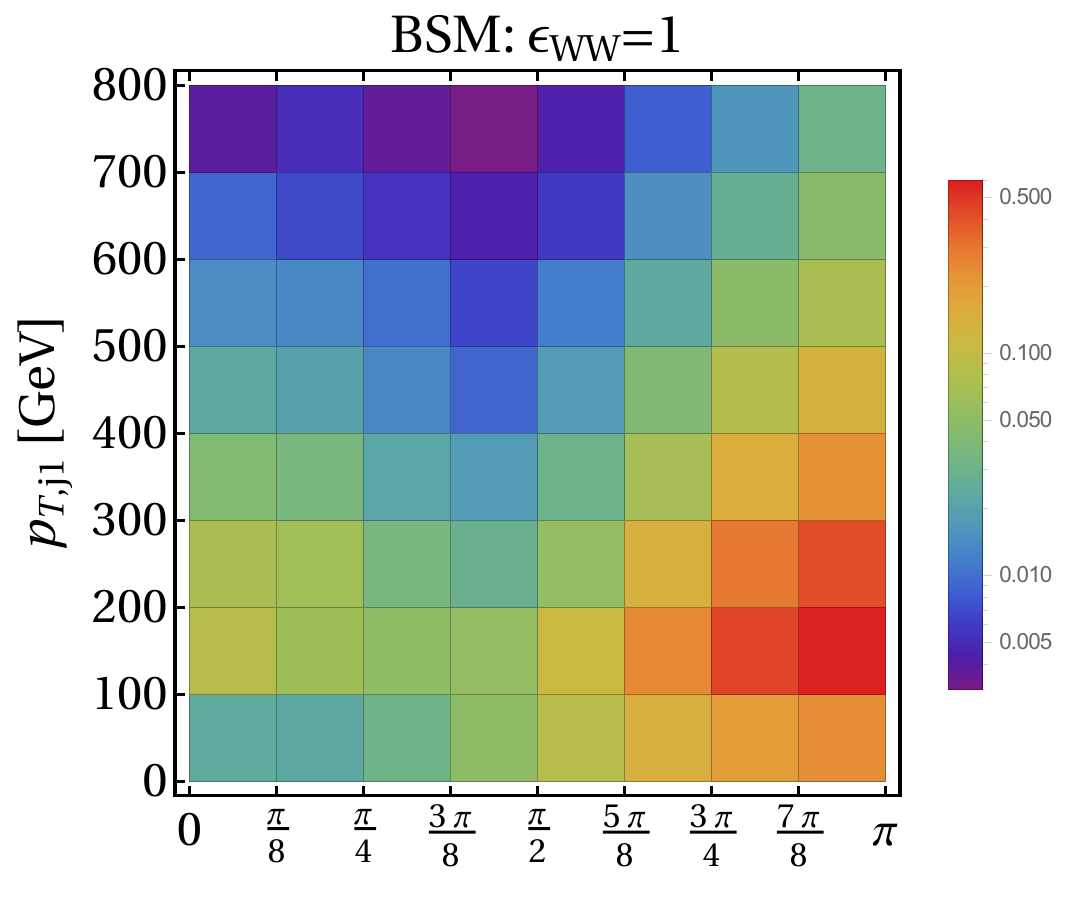}}\\ \vspace{-0.48cm}
  \subfloat{\includegraphics[width=0.4\textwidth]{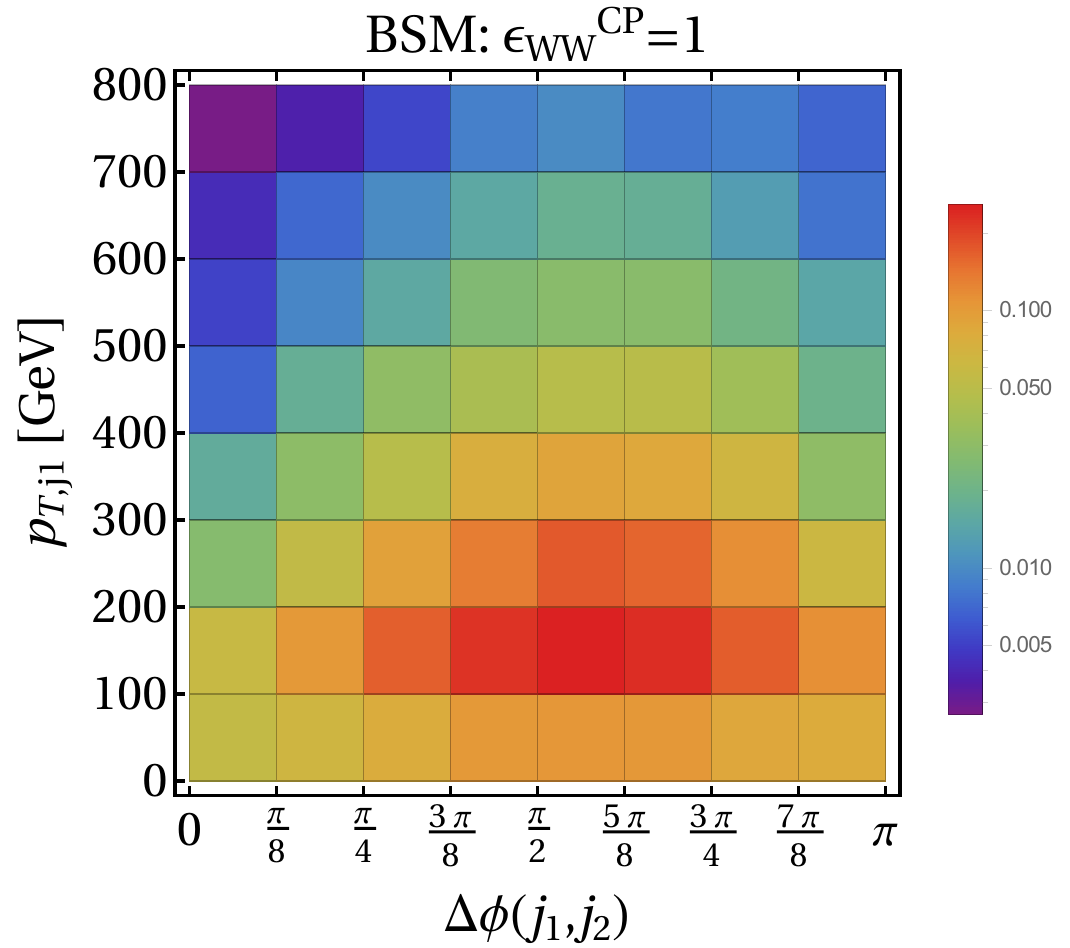}}
    \caption{NLO+PS predictions of cross section (pb)  for $\Delta \phi({\rm j}_1, {\rm j}_2)$-$p_{\rm T, j_1}$ distributions in VBF Higgs production at 13 TeV.  2D distributions are shown for the SM and PO extensions with $\epsilon_{WW}=1$, $\epsilon_{WW}^{CP}=1$.
    \label{VBF_2D}}
\end{figure}

Events for VBF Higgs production, without $V(\to q\bar q)H$ contributions, are generated via\footnote{In order to include also the $V(\to q\bar q)H$ contribution one should use instead {\tt > generate p p > h j j  HPO=1 QED=2 [QCD]}.}:
\begin{verbatim}
> generate p p > h j j $$ a z w+ w- \
                    HPO=1 QED=2 [QCD]
\end{verbatim}
Formally this processes comprises IR-divergent pentagon one-loop diagrams with two massive vector boson propagators. In the available automated frameworks such contributions can typically only be evaluated in Feynman gauge. Thus, in line with the default of  \MadGraph such diagrams should be discarded in conjunction with the \HPO model. This is motivated by the tiny 
numerical impact of these diagrams in the VBF phase-space. However, this introduces a formal mismatch of the IR poles in the computation. This is a well known issue for VFB Higgs production in \MadGraph and a pragmatic solution requires to turn off the check for IR pole cancellation. 
This can be achieved by modifying the {\tt  {\#}IRPoleCheckThreshold} parameter in the '{\tt FKS{\_}params.dat}' file from {\tt 1.0d-5} to {\tt -1.0d0}.

In our numerical analysis we  employ the same VBF cuts as in Ref.~\cite{Greljo:2015sla}, i.e. we require
\be
	p_{{\rm T, j}_{1,2}} > 30 \GeV~, \quad
	|\eta_{{\rm j}_{1,2}}| < 4.5~, \quad
	m_{{\rm j}_1 {\rm j}_2} > 500 \GeV~,
\ee
where jets are ordered according to their $p_T$.
In Fig.~\ref{VBF-plot} we show the NLO (solid lines) and LO (dashed lines) leading-jet $p_T$ distribution for the benchmark points listed in Table~\ref{tab:bench}. Again, in the lower panel the ratio of the NLO over LO predictions is presented, indicating a rather flat and universal $K$-factor across a large kinematic regime.

At particle-level more than two jets can be reconstructed in VBF events. Thus, different jet definitions and VBF tagging techniques can have relevant impact on kinematic distributions in VBF Higgs production \cite{Rauch:2017cfu}. In particular, it might be interesting to optimize these choices such that the sensitivity to the different PO is maximised. We leave this to future studies.

Finally, in Fig.~\ref{VBF_2D} we illustrate the NLO 2D distributions of the azimuthal angle difference of two leading jets $\Delta \phi ({\rm j}_1, {\rm j}_2)$ and the leading jet transverse momentum $p_{T,j1}$ in VBF Higgs production. These 2D distributions are sensitive to the presence of transversal ($\epsilon_{ZZ,WW}$) and CP-odd ($\epsilon_{ZZ,WW}^{CP}$) POs, since these are characterized by a different Lorentz structure than the SM $hVV$ coupling.  In order to illustrate such effects we present the distributions for two rather extreme PO benchmark points ($\epsilon_{WW}=1$ and $\epsilon_{WW}^{CP}=1$), alongside the SM prediction. We also checked that taking such 2D distributions into account in a global PO fit will substantially improve the sensitivity to the transversal and CP-odd PO, compared to using only the leading jet $p_T$ distribution.

\section{PO-dependence of the Simplified Template cross sections at NLO in QCD}
\label{sec:STXS}

As an example of a practical application of the \HPO tool, we compute the \emph{simplified template cross sections} (STXS) for $Zh$ production as a function of the Higgs PO at NLO in QCD.
The STXS, introduced in Chapter III.2 of Ref.~\cite{deFlorian:2016spz}, are cross sections defined in some simple and idealized bins. The choice of such bins is motivated by optimizing the sensitivity to BSM effects while minimizing the necessary acceptance corrections (thus theory dependence) by choosing simple selection cuts to a phase-space close to the realistic fiducial region. Several stages with more and more granular bins are envisaged with increasing luminosity.

In the case of $Zh$ production, with $Z \to \ell^+ \ell^-$ or $\bar{\nu}\nu$ decays, the kinematical variable chosen for the binning is the $p_T$ of the $Z$ boson ($E_T^{\rm miss}$ in the case of neutrinos), with bins $[0 - 150 - 250 - 400 - \infty]$ GeV. Furthermore, a possible split of the $[150 - 250]$ GeV bin is envisaged in 0-jet and $\geq$1-jet categories. For simplicity of this preliminary analysis we neglect this jet categorization.
An overall selection on the Higgs rapidity $|y_h| < 2.5$ is applied.

The idea behind the STXS is that, on the one hand, the experimental collaborations could provide a measurement of such cross sections in a very model-independent way by combining Higgs data in different channels. On the other hand, since the STXS are defined at the reconstruction level with simple kinematical cuts, they can easily be computed in a given BSM model in order to derive the limits.
The Higgs PO-dependence of the STXS $p_T$ bins in $Zh$ production, for example, can be used to obtain global fits in the PO formalism.

The $p p \to Z h$ events are generated in the same way as described in Section~\ref{sec:results} and subsequently analyzed in order to apply the selection cut on $y_h$ described above and separate them into different $p_T^Z$ bins. In this way we obtain the cross section in each STXS bin as a quadratic function of the Higgs PO. The result, normalised to the SM value, is:
\begin{align}
	\!\! R_{0-150} =&
	k_{ZZ}^2 + k_{ZZ} (27 \epsilon_{Z u_L} - 28 \epsilon_{Z d_L} - 14 \epsilon_{Z u_R} + 4.7 \epsilon_{Z d_R}) \nonumber \\
	& + 10^2 [6.0 (\epsilon_{Z u_L}^2 + \epsilon_{Z u_R}^2) + 4.7 (\epsilon_{Z d_L}^2 +  \epsilon_{Z d_R}^2)]  ~, \nonumber \\
	\!\! R_{150-250} =&
	k_{ZZ}^2 + k_{ZZ} (80 \epsilon_{Z u_L} - 71 \epsilon_{Z d_L} - 37 \epsilon_{Z u_R} + 14 \epsilon_{Z d_R}) \nonumber \\
	& + 10^3 [4.5 (\epsilon_{Z u_L}^2 + \epsilon_{Z u_R}^2) + 3.2 (\epsilon_{Z d_L}^2 +  \epsilon_{Z d_R}^2)]  ~, \nonumber \\
	\!\! R_{250-400} =&
	k_{ZZ}^2 + k_{ZZ} (180 \epsilon_{Z u_L} - 145 \epsilon_{Z d_L} - 83 \epsilon_{Z u_R} + 31 \epsilon_{Z d_R}) \nonumber \\
	& + 10^4 [2.2 (\epsilon_{Z u_L}^2 + \epsilon_{Z u_R}^2) + 1.5 (\epsilon_{Z d_L}^2 +  \epsilon_{Z d_R}^2)] ~, \nonumber \\
	\!\! R_{400-\infty} =&
	k_{ZZ}^2 + k_{ZZ} (610 \epsilon_{Z u_L} - 340 \epsilon_{Z d_L} - 235 \epsilon_{Z u_R} + 145 \epsilon_{Z d_R}) \nonumber \\
	& + 10^5 [2.5 (\epsilon_{Z u_L}^2 + \epsilon_{Z u_R}^2) + 1.4 (\epsilon_{Z d_L}^2 +  \epsilon_{Z d_R}^2)] ~,
\end{align}
where $R_x \equiv \sigma_i / \sigma^{\rm SM}_i$ in each STXS bin.

\section{Conclusions}
\label{conclusions}

In this work we {have presented} an implementation of the Higgs Pseudo-Observables framework for electroweak Higgs production in the \HPO \,\UFO model for Monte Carlo event generation at NLO in QCD.
The extension of the previous version of the tool to NLO accuracy {will allow to perform a robust interpretation of high-statistics data on 
 $Vh$ and VBF Higgs production in this general BSM formalism.

{The tool takes into account  all PO} relevant for EW Higgs production processes, under the hypothesis of an $U(2)^3$ flavour symmetry. This symmetry reduces the number of independent $Vhqq'$ quark contact terms to six, or to five if one further assumes CP conservation. On top of this, all the flavour-universal PO and leptonic contact terms (relevant for leptonic Higgs decays) are also included. All these parameters can be easily accessed and modified for example from the {\tt param{\_}card.dat} in \MadGraph.

In order to illustrate the use of the \HPO model we {have presented} simulations of $Zh$, $Wh$, and VBF Higgs production for a set of benchmark points. For each benchmark point we {have studied} the differential cross section in a relevant kinematical variable, and the effect of the NLO corrections. {Our analysis shows that}, 
for a suitable choice of factorization and renormalization scales such as $H_T/2$, the NLO corrections are relevant, $\mathcal{O}(20\%)$, but rather flat in the kinematical regime of interest. This suggests a fair convergence of the perturbative expansion.
In the case of VBF production we {have shown} how the correlation between the leading jet $p_T$ and the azimuthal angle difference of the two leading jets, $\Delta \phi (j_1, j_2)$, is a very sensitive probe of coupling structures different from the SM ones.
Finally, we {have presented} the first computation of simplified template cross sections for $Zh$ production as a function of Higgs pseudo-observables at NLO accuracy.

{As is well known,} higher-order EW corrections in EW Higgs production become very important at large energies due to the appearance of Sudakov logarithms. The dominant EW Sudakov logarithms are universal and can be factorized from the hard part of the scattering process \cite{Denner:2001gw}. Nonetheless, being of electroweak nature, these corrections are expected to generate {a non-trivial} mixing among different PO. {This implies that their implementation in the PO framework, especially 
in the context of a Monte Carlo event generator, is a non-trivial task~\cite{IsidoriPattori}. While waiting for the completion of this program, we stress that 
current data on EW Higgs production are dominated by events near the kinematical threshold, where the effect of EW corrections is not log-enhanced.\footnote{At present no general-purpose Monte Carlo event generator includes such EW corrections for BSM models.}}

To sum up, the \HPO  ~ --- event simulation tool for Higgs pseudo-observables framework --- is upgraded to NLO in QCD and is publicly available at\\
\centerline{\url{http://www.physik.uzh.ch/data/HiggsPO}\,\,.} 

\begin{acknowledgements}
This research was supported in part by the Swiss National Science Foundation (SNF) under contract 200021-159720.
\end{acknowledgements}

\bibliographystyle{JHEP}
\bibliography{paper}

\end{document}